\documentclass[iop]{emulateapj}

\usepackage[tight]{subfigure}
\usepackage{amsmath}
\usepackage{graphicx}
\usepackage{color}
\usepackage{verbatim}

\newcommand{\gccm}{\ensuremath{\mathrm{g} \, \mathrm{cm}^{-3}}}
\newcommand{\kms}{\ensuremath{\mathrm{km} \, \mathrm{s}^{-1}}}
\newcommand{\cms}{\ensuremath{\mathrm{cm} \, \mathrm{s}^{-1}}}
\newcommand{\snia}{SN~Ia}
\newcommand{\sneia}{SNe~Ia}
\newcommand{\snfe}{SN~2011fe}

\shorttitle{Constraining SN~Ia models with observations of SN~2011fe} 
\shortauthors{F. K. R\"opke et al.}

\begin{document}

\title{Constraining Type I\lowercase{a} supernova models: SN~2011\lowercase{fe} as a test case}

\author
{
    F.~K.~R\"opke,\altaffilmark{1,2}
    M.~Kromer,\altaffilmark{2}
    I.~R. Seitenzahl,\altaffilmark{1,2}
    R.~Pakmor,\altaffilmark{3}
    S.~A.~Sim,\altaffilmark{4}
    S.~Taubenberger,\altaffilmark{2}
    F.~Ciaraldi-Schoolmann,\altaffilmark{2}
    W.~Hillebrandt,\altaffilmark{2}
    G.~Aldering,\altaffilmark{5}
    P.~Antilogus,\altaffilmark{6}
    C.~Baltay,\altaffilmark{7}
    S.~Benitez-Herrera,\altaffilmark{2}
    S.~Bongard,\altaffilmark{6}
    C.~Buton,\altaffilmark{8}
    A.~Canto,\altaffilmark{6}
    F.~Cellier-Holzem,\altaffilmark{6}
    M.~Childress,\altaffilmark{5,9}
    N.~Chotard,\altaffilmark{10}
    Y.~Copin,\altaffilmark{10}
    H.~K. Fakhouri,\altaffilmark{5,9}
    M.~Fink,\altaffilmark{1,2}
    D.~Fouchez,\altaffilmark{13}
    E.~Gangler,\altaffilmark{10}
    J.~Guy,\altaffilmark{6} 
    S.~Hachinger,\altaffilmark{2}
    E.~Y. Hsiao,\altaffilmark{5}
    C.~Juncheng,\altaffilmark{14}
    M.~Kerschhaggl,\altaffilmark{8}
    M.~Kowalski,\altaffilmark{8}
    P.~Nugent,\altaffilmark{11}
    K.~Paech,\altaffilmark{8}
    R.~Pain,\altaffilmark{6}
    E.~Pecontal,\altaffilmark{12}
    R.~Pereira,\altaffilmark{10}
    S.~Perlmutter,\altaffilmark{5,9}
    D.~Rabinowitz,\altaffilmark{7}
    M.~Rigault,\altaffilmark{10}  
    K.~Runge,\altaffilmark{5}
    C.~Saunders,\altaffilmark{5,9}
    G.~Smadja,\altaffilmark{10}
    N.~Suzuki,\altaffilmark{5}
    C.~Tao,\altaffilmark{13,14}
    R.~C. Thomas,\altaffilmark{11}
    A.~Tilquin,\altaffilmark{13}
    C.~Wu\altaffilmark{6,15}
}

\altaffiltext{1}
{
  Institut f{\"u}r Theoretische Physik und Astrophysik, 
  Universit{\"at} W{\"u}rzburg, 
  Am Hubland, 
  D-97074 W{\"u}rzburg, Germany
} 
\altaffiltext{2}
{
  Max-Planck-Institut f\"ur Astrophysik,
  Karl-Schwarzschild-Str. 1, 
  D-85741 Garching, Germany
} 

\altaffiltext{3}
{
  Heidelberger Institut f\"{u}r Theoretische Studien, 
  Schloss-Wolfsbrunnenweg 35, 
  69118 Heidelberg, Germany
}
\altaffiltext{4}
{
    Research School of Astronomy and Astrophysics,
    The Australian National University,
    Mount Stromlo Observatory,
    Cotter Road, Weston Creek ACT 2611 Australia
}
\altaffiltext{5}
{
    Physics Division, Lawrence Berkeley National Laboratory, 
    1 Cyclotron Road, Berkeley, CA, 94720
}
\altaffiltext{6}
{
    Laboratoire de Physique Nucl\'eaire et des Hautes \'Energies,
    Universit\'e Pierre et Marie Curie Paris 6, Universit\'e Paris Diderot Paris 7, CNRS-IN2P3, 
    4 place Jussieu, 75252 Paris Cedex 05, France
}
\altaffiltext{7}
{
    Department of Physics, Yale University, 
    New Haven, CT, 06250-8121
}
\altaffiltext{8}
{
    Physikalisches Institut, Universit\"at Bonn,
    Nu\ss allee 12, 53115 Bonn, Germany
}
\altaffiltext{9}
{
    Department of Physics, University of California Berkeley,
    366 LeConte Hall MC 7300, Berkeley, CA, 94720-7300
}
\altaffiltext{10}
{
    Universit\'e de Lyon, F-69622, Lyon, France ; Universit\'e de Lyon 1, Villeurbanne ; 
    CNRS/IN2P3, Institut de Physique Nucl\'eaire de Lyon.
}
\altaffiltext{11}
{
    Computational Cosmology Center, Computational Research Division, Lawrence Berkeley National Laboratory, 
    1 Cyclotron Road MS 50B-4206, Berkeley, CA, 94611
}
\altaffiltext{12}
{
    Centre de Recherche Astronomique de Lyon, Universit\'e Lyon 1,
    9 Avenue Charles Andr\'e, 69561 Saint Genis Laval Cedex, France
}
\altaffiltext{13}
{
    Centre de Physique des Particules de Marseille , 163, avenue de Luminy - Case 902 - 13288 Marseille Cedex 09, France
}
\altaffiltext{14}
{
    Tsinghua Center for Astrophysics, Tsinghua University, Beijing 100084, China 
}
\altaffiltext{15}
{
    National Astronomical Observatories, Chinese Academy of Sciences, Beijing 100012, China
}

\begin{abstract}
  The nearby supernova \snfe\ can be observed in unprecedented
  detail. Therefore, it is an important test case for Type Ia
  supernova (SN~Ia) models, which may bring us closer to understanding
  the physical nature of these objects. Here, we explore how available
  and expected future observations of \snfe\ can be used to
  constrain SN~Ia explosion scenarios. We base our discussion on
  three-dimensional simulations of a delayed detonation in a
  Chandrasekhar-mass white dwarf and of a violent merger of two white
  dwarfs---realizations of explosion models appropriate for two of the
  most widely-discussed progenitor channels that may give rise to
  SNe~Ia. Although both models have their shortcomings in reproducing
  details of the early and near-maximum spectra of \snfe\
  obtained by the Nearby Supernova Factory (SNfactory), the overall
  match with the observations is reasonable. The level of agreement is
  slightly better for the merger, in particular around maximum, 
  but a clear preference for one model over the other is
  still not justified. Observations at late epochs, however, hold 
  promise for discriminating the explosion scenarios in a straightforward
  way, as a nucleosynthesis effect leads to differences in the $^{55}$Co 
  production. \snfe\ is close enough to be followed sufficiently 
  long to study this effect.
\end{abstract}


\section{Introduction}\label{intro}Perhaps the most fundamental
problem hindering a better understanding of \snia\ explosions is the
unclear nature of the progenitor system.
One way of addressing this problem is to carry out numerical
simulations for different scenarios that involve thermonuclear
explosions of white dwarfs (WDs) and to compare the results with
observations. Obviously, detailed observational data are a
prerequisite for this approach.  At the same time, the comparison
should be based on models that avoid free parameters in the
description of the explosion mechanism, as far as possible. Only then
the predictive power of theoretical models will be sufficient to
discriminate between explosion models and to draw conclusions about
progenitor systems.

In addition to the possibility of directly constraining the progenitor
system from archival data \citep{li2011b,liu2011a} or early
observations \citep{brown2011a,nugent2011a, bloom2012a}, the recently
discovered nearby \snia\ 2011fe offers a unique opportunity for a
comparison with explosion models.  Of particular value are the
possibility to follow this close object photometrically to extremely
late epochs and the exact knowledge of the explosion time.

\snfe\ was first detected by the Palomar Transient Factory on 2011
August 24.167 in M101 \citep{nugent2011b}.  A preliminary analysis of
our data indicates that it reached an apparent $B$-band peak magnitude
of $9.9$ on September 11. Combined with the derived explosion date of
2011 August 23.7 \citep{nugent2011a}, the $B$-band rise time of \snfe\
is $\sim$$18.3\,\mathrm{d}$---a typical value for normal \sneia\
\citep{conley2006a,hayden2010a}. Assuming a distance to M101 of
$6.4\,\mathrm{Mpc}$ \citep{shappee2011a}, \snfe\ is a normal \snia\
with $M_{B,\mathrm{max}}=-19.13$, having produced $\sim$$0.6 \,
M_\odot$ of $^{56}$Ni \citep{stritzinger2006a}. The identification of
\snfe\ as a prototypical \snia\ is also corroborated by the observed
spectra (as shown below).

With the development of three-dimensional simulations of thermonuclear
explosions in carbon--oxygen WDs and of the subsequent radiative
transfer (RT) leading to the formation of the observables, a new
generation of models is currently becoming available. These have the
advantage that the explosion physics is represented in a far less
parameterized manner than in previous one-dimensional models. Due to
their improved predictive power, a comparison with observational data
would in principle allow us to constrain the explosion scenario of
\sneia. However, no currently available multi-dimensional model
reaches the level of agreement that can be obtained fitting
one-dimensional semi-empirical models to data. This challenges the
interpretation of the comparison between the new models and \snia\
data.

Here, we address the question of whether \snfe\ can be explained by
models of an exploding Chandrasekhar-mass WD (realized as a delayed
detonation) or a violent merger of two WDs. These scenarios can lead
to observables that resemble normal \sneia\ \citep{mazzali2007a,
  kasen2009a, pakmor2012a}, but they differ fundamentally in the
explosion mechanism, the mass and the structure of the ejecta. A
discrimination between them based on comparison with observations
would help to shed light on the open question of the progenitor
system. An explosion of the WD near the Chandrasekhar mass is usually
attributed to the single-degenerate progenitor channel in which a
carbon-oxygen WD accretes matter from a non-degenerate companion;
however, the formation of a Chandrasekhar-mass object in a merger of
two WDs cannot be excluded.  Our second scenario results from the
merger of two WDs with similar and rather high masses adding up to a
total of $2\,M_\odot$. Both models are set up to produce
$\sim$$0.6\,M_\odot$ of $^{56}$Ni but apart from that they follow
generic assumptions and are not tuned to fit the data of \snfe.

\section{Explosion Models}The most promising way of producing
observables in reasonable agreement with observations of normal
\sneia\ from an explosion of a Chandrasekhar-mass WD is the delayed
detonation mechanism \citep{khokhlov1991a}. We model this scenario
using the techniques described by \citet{reinecke1999a,roepke2005b,
  schmidt2006c} and \citet{roepke2007b}.

An isothermal ($T=5\times10^5\,\mathrm{K}$) WD composed of carbon and
oxygen in equal parts by mass was set up in hydrostatic equilibrium
with a central density of $2.9\times10^9\,\gccm$ and an electron
fraction of $Y_e=0.498864$, corresponding to solar metallicity. The
model was discretized on a three-dimensional Cartesian moving grid
\citep{roepke2005c} with $512^3$ cells consisting of two nested parts.
To reach the intended $^{56}$Ni production, the initial deflagration
was ignited in $100$ sparks placed randomly in a Gaussian distribution
within a radius of $150\,\mathrm{km}$ from the WD's center on the
inner grid, which had a resolution of $1.92\times10^5\,\mathrm{cm}$.
After an initial deflagration phase similar to that described by
\citet{roepke2007c}, a detonation was triggered at every location on
the flame where the fuel density was in the range of $6$ to
$7\times10^6 \,\gccm$ and the grid cell contained preferentially fuel
material, provided that the turbulent velocity fluctuations exceeded
$10^8\,\cms$ at a significant fraction of the flame area and persisted
for sufficiently long times. This loosely follows the criteria proposed
by \citet{woosley2009a}. Since the initiation of a detonation
proceeds on scales that are not resolved in our simulations, the probability of finding high turbulent
velocities on unresolved scales is extrapolated applying the procedure
of \citet{roepke2007d}. The evolution was followed to a time of
$100\,\mathrm{s}$ after ignition, by which homologous expansion of the
ejecta was reached to a good approximation. This model, called N100,
is part of a larger set of delayed-detonation simulations (Seitenzahl
et al., in preparation).

The details of the nucleosynthesis in this explosion were determined
from thermodynamic trajectories recorded by $10^6$ tracer particles
distributed in the exploding WD \citep{travaglio2004a,
  seitenzahl2010a}. The characteristics of the model are summarized in
Table~\ref{tab:models}.

The second simulation we discuss models the inspiral, merger and
explosion of two WDs with $1.1\,M_\odot$ and $0.9\,M_\odot$,
respectively. Details of the corresponding simulations are given by
\citet{pakmor2012a}. While the inspiral and merger phases were
followed with a version of the SPH code \textsc{Gadget}
\citep{springel2005a}, the subsequent thermonuclear detonation was
modeled with techniques similar to those employed in
N100. The question of whether a detonation triggers at the
  interface between the two merging stars is
  controversial. Simulations with sufficiently high numbers of SPH
  particles, such as presented here, show the formation of a hot spot,
  and we assume a detonation to trigger at this location when the
  temperature exceeds $2.5\times 10^{9}\,\mathrm{K}$ in material of $\rho\approx
  2\times 10^{6}\,\gccm$, relying on
  the microscopic simulations of detonation initiation by
  \citet{seitenzahl2009b}.  Again, the evolution was followed up to
$100\,\mathrm{s}$ and the composition of the ejecta was
determined in a post-processing step. The results of this simulation are given
in Table~\ref{tab:models}.

\begin{figure*}
  \centerline{
  \includegraphics[width=\linewidth]{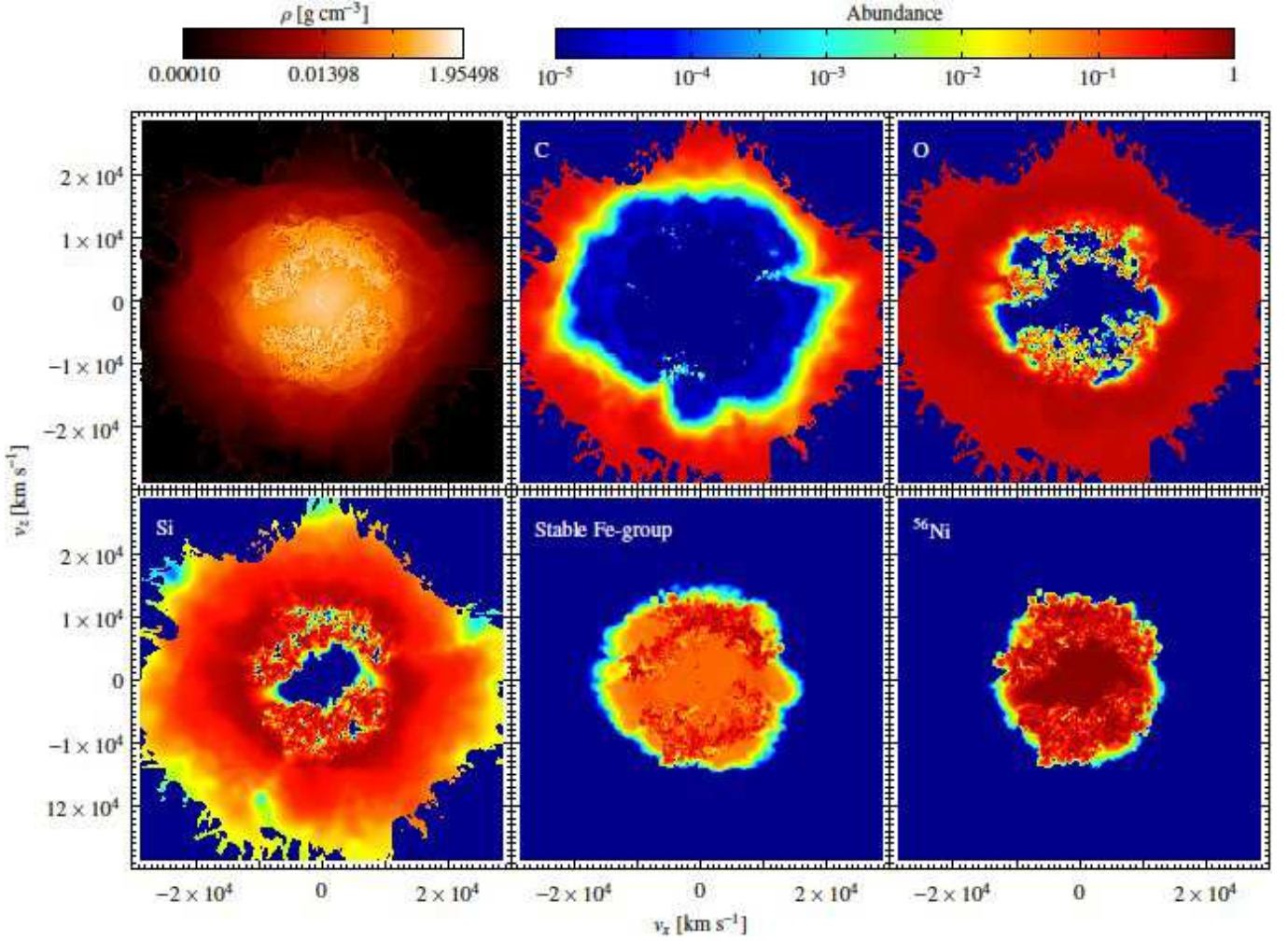}}
\caption{Slices through our delayed-detonation model N100 in the
  $x$--$z$-plane showing the density (top left) and abundance
  distribution of selected species at 100\,s after explosion.\label{fig:deldet_model}}
\end{figure*}

\begin{figure*}
  \centerline{
  \includegraphics[width=\linewidth]{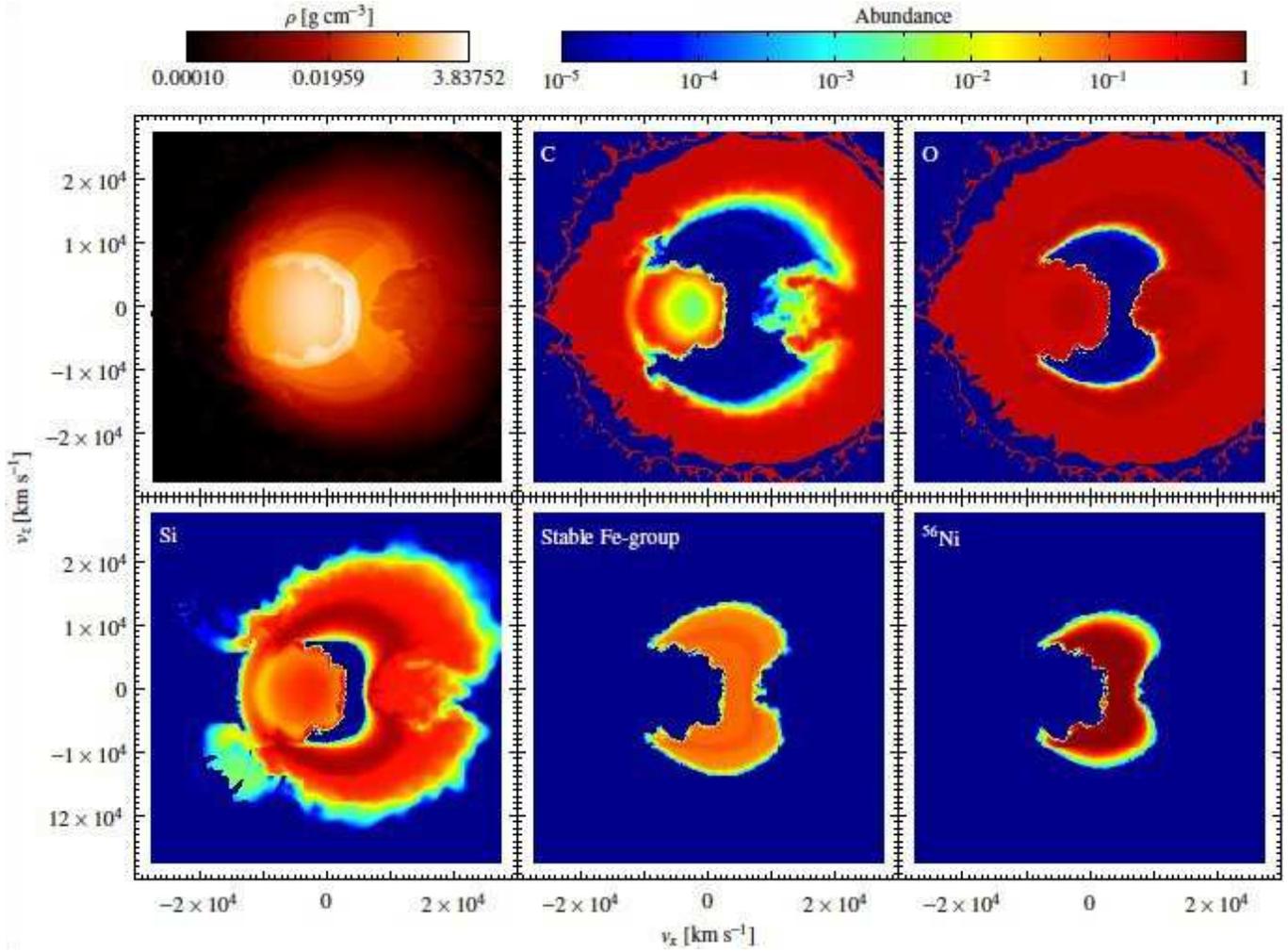}}
\caption{Same as Fig.~\ref{fig:deldet_model} but for the merger
  model.\label{fig:merger_model} }
\end{figure*}

Density and composition of both models in homologous expansion are
visualized in Figs.~\ref{fig:deldet_model} and
\ref{fig:merger_model}. We note that the ejecta structure resulting
from the WD-WD merger differs fundamentally from that of
N100. This is because the explosion of the secondary WD happens
shortly after that of the primary. Therefore, the outer ejecta
material originates from the primary. At the onset of the explosion,
the primary had a radius below $0.02\,R_\odot$ making our \emph{violent}
merger scenario consistent with the constraint on the radius of the
exploding object derived by \citet{bloom2012a}.

\begin{table*}
\begin{center}
\caption{Model characteristics.}
\label{tab:models}
\hspace{-2.5cm}
\begin{tabular}{lll} \hline
 & delayed detonation (N100) & violent merger\\
 \hline
total ejecta mass [$M_\odot$] &1.40 &
1.95\footnote{$0.05\,\mathrm{M_\odot}$ are lost during the explosion simulation because of
the finite extent of the grid} \\
asymptotic kinetic energy of ejecta [$10^{51}\,\mathrm{erg}$] & 1.45 & 1.7 \\ \hline
$^{56}$Ni mass [$M_\odot$] &0.604& 0.616\\
total iron group [$M_\odot$] & 0.839 & 0.697 \\
total intermediate mass elements [$M_\odot$] &0.454& 0.5 \\
carbon mass [$M_\odot$] &0.003& 0.153\\
oxygen mass [$M_\odot$] &0.101& 0.492 \\
combined mass of $^{55}$Fe and $^{55}$Co [$M_\odot$] & $1.33\times10^{-2}$ & $3.73\times10^{-3}$ \\
combined mass of $^{57}$Ni and $^{57}$Co [$M_\odot$] &$1.88\times10^{-2}$&$1.49\times10^{-2}$\\ \hline
$B$-band rise time [days] &         16.6 &  20.8 \\
$B$-band peak luminosity [mag] &  $-19.0$ & $-19.0$ \\
$\Delta m_{15}(B)$ [mag]  &  1.34 &  0.95 \\ \hline
$\mathrm{D}_\mathrm{late}^{500}\equiv m_{1400\mathrm{d}} - m_{900\mathrm{d}}$ in leptonic light curve [mag] & 2.25 & 2.65 \\ 
$\mathrm{D}_\mathrm{late}^{1000}\equiv m_{1900\mathrm{d}} - m_{900\mathrm{d}}$ in leptonic light curve [mag] & 3.20 & 3.87 \\
\end{tabular}
\end{center}
\end{table*}

\section{Comparison with spectra of SN~2011\lowercase{fe}}From the
nucleosynthesis tracer particles we constructed detailed abundance
distributions of the explosion ejecta at 100\,s
and mapped them to $50^3$ Cartesian grids. These grids
were then used to derive synthetic light curves and spectra with the 
Monte Carlo RT code \textsc{Artis}
\citep{kromer2009a,sim2007b}.  To this end, we simulated the
propagation of $10^8$ photon packets from 2 to 120 days after
explosion using the cd23\_gf-5 atomic dataset of \citet{kromer2009a},
which is based on the lines contained in the CD23 compilation of
\citet{kurucz1995a}. To account for higher ionization at early times,
we added the ionization stages \textsc{vi} and \textsc{vii}
for Sc to Ni, leading to a total of ${\sim}5\times10^5$ atomic lines.

Both our models yield a $B$-band peak magnitude of $-19.0$, roughly in
agreement with that observed for \snfe. Their rise times, however,
differ: while N100 reaches $B$-band maximum after 16.6\,d, the
merger takes 20.8\,d (further parameters of our synthetic light
curves are given in Table~\ref{tab:models}). Thus, neither of the
models gives a perfect match to the light curves of \snfe\ but both
are sufficiently close to warrant further investigation.

\begin{figure*}
  \centerline{\includegraphics[width=\linewidth]{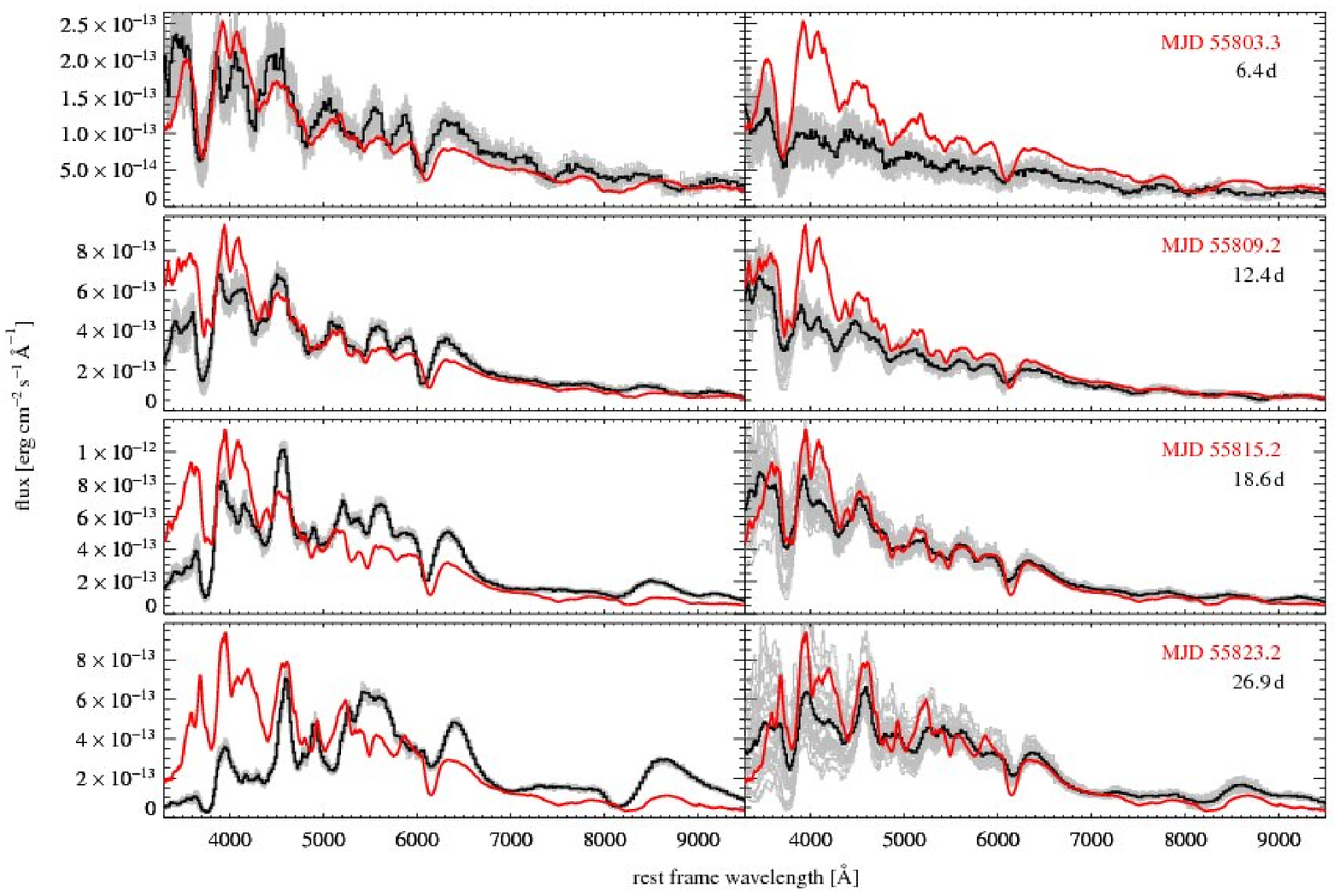}}
  \vspace*{4ex}
  \centerline{\includegraphics[width=0.995\linewidth]{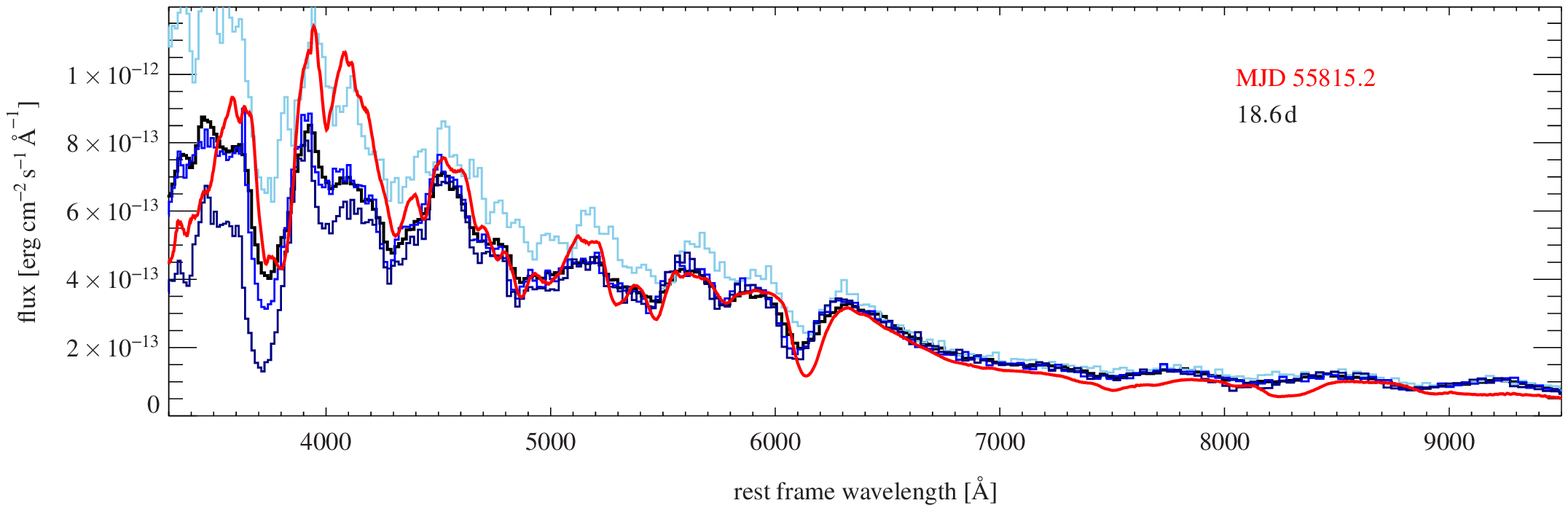}}

  \caption{Spectral evolution of our delayed detonation N100 (left)
    and merger model (right) from 6 to 27 days after the
    explosion. The angle-averaged spectrum is plotted in black while
    25 spectra for representative viewing angles are shown in
    gray (the variability with viewing angle of the earliest spectra 
    is dominated by Monte Carlo noise in both models). For comparison 
    the observed spectra of \snfe\ are
    over-plotted in red assuming an explosion date at August 23.7 (MJD
    55796.7; \citealt{nugent2011a}). The observations were corrected
    for Galactic reddening assuming $E(B-V)_\mathrm{Gal}=0.009$\,mag
    \citep{schlegel1998a} and de-redshifted according to a
    heliocentric radial velocity $v_\mathrm{hel}=241\,\kms$ given by
    \citet{devaucouleurs1991a}. Reddening from the host is negligible. The bottom panel compares the observed
    spectrum of \snfe\ near $B$-band maximum (red) with
    synthetic spectra from the merger corresponding to three
    different viewing angles (blue colors) and the angle-average
    (black).\label{fig:spectra}}
\end{figure*}

In Fig.~\ref{fig:spectra} we compare synthetic spectra from our models
with flux-calibrated spectra of \snfe\ taken by the SNfactory
collaboration with the SNIFS instrument \citep{aldering2002a} on the
University of Hawaii 2.2\,m telescope on Mauna Kea. To our knowledge
this is the first direct comparison of consistent three-dimensional
\snia\ models with a spectrophotometric time-series. Overall, the
spectra of both explosion scenarios reproduce the main features of the
observed spectra and the flux level reasonably well (note that these
are not fits but predictions from ``first-principle'' models and that
\emph{absolute} fluxes are compared). In detail, however, there are
problems in both models.

The predicted absorption features are blue-shifted with respect to the
observations. Although this effect is stronger in N100, it is
also visible for the merger, indicating too high ejecta velocities in
the models. Since the mass of the exploding object is very different
in the two cases, the nuclear energy release is likely too high in both
explosion processes. A potential way to cure this problem is to
increase the oxygen abundance in the progenitor WDs at the expense of
carbon thus increasing the average nuclear binding energy of the fuel.

While N100 is only marginally too bright at the early epochs,
the merger is clearly too faint. This corresponds to the
shorter/longer rise time of the respective model (see
Tab.~\ref{tab:models}) compared to \snfe. Around $B$-band maximum at
${\sim}18$\,d the merger compares favorably to the observed
spectra. The flux level and the overall shape of its synthetic
spectra match the data better than those of N100. After maximum
the agreement with the observations deteriorates for both
models, although the effect is more drastic for N100. In
particular, the models fail to reproduce the
spectral features between $5000\,$\AA\ and
$6000\,$\AA\@. Moreover, both models become
redder faster than the observation. Again, this trend is more
pronounced in N100, but is also visible in the merger.

Figs.~\ref{fig:deldet_model} and \ref{fig:merger_model} show that
iron-group elements (IGEs) extend significantly beyond velocities of
$10,000\,\kms$ in N100 but also in some directions in the merger. The
W7 model of \citet{nomoto1984a}, which is known to reproduce
observables of \sneia\ well, does not contain IGEs at such high
velocities. However, they are reported in abundance tomographies of
the normal SNe 2002bo \citep{stehle2005a} and 2004eo
\citep{mazzali2008b} and \citet{nugent2011a} identify iron in the
earliest spectra of \snfe\ at velocities of $16,000\,\kms$.  Our
synthetic spectra do not show mismatches with observed lines that can
be directly attributed to high-velocity IGEs. It is possible, however,
that they contribute to the fast reddening of the models. As for the
high ejecta velocities, a decreased carbon/oxygen ratio in the
exploding WDs may alleviate this problem.

An important difference between the two models is also visible from
Figs.~\ref{fig:deldet_model} and \ref{fig:merger_model}: N100
is---apart from small-scale anisotropies---roughly spherical (see also
\citealt{blondin2011a}). In contrast, the merger shows pronounced
large-scale asymmetries. This is reflected in a strong viewing-angle
dependence in its spectra which increases after maximum due to the
growing asymmetry of the inner ejecta for smaller radii. As
demonstrated in the lower plot of Fig.~\ref{fig:spectra} individual
line-of-sight spectra from the merger reproduce the observation at
least as well as the angle-average. Nevertheless, the high level of
asymmetry in the merger could be in conflict with the observed
spectral homogeneity of normal \sneia\ and the low level of continuum
polarization observed \citep{wang2008a}.  Currently our RT simulations
do not include polarization.  Note, however, that \citet{smith2011a}
report a continuum polarization of 0.2--0.4\% for \snfe\ which they
interpret as a sign of persistent asymmetry in the last-scattering
surface.

The fact that the merger reproduces the observed spectra better
than N100 at maximum light (and later) suggests that the chemical
structure of its deeper ejecta, which dominate the spectrum formation
at this epoch, is closer to that of \snfe. Since neither of the models
matches the optical data of early epochs perfectly, our comparison
gives slight preference to a WD merger scenario over a delayed
detonation in a Chandrasekhar-mass WD as an explanation for this
object. But as there are major shortcomings in both models, a
definitive conclusion cannot be drawn. Observables other than
maximum-light spectra may, however, have more discriminating power. 
We discuss promising possibilities in the following.

\section{Late-time observables}While optical data taken before and
around peak brightness probe predominantly the outermost layers,
observations at later epochs are sensitive to the core of the 
ejecta. Since this is the region where the differences between the two
models considered here are most pronounced, late-time observables are
a very useful diagnostic tool.

A fundamental difference between our two explosion scenarios is the
density at which the material is burned in the thermonuclear
combustion.  Due to the high central density of the Chandrasekhar-mass
WD, substantial burning proceeds on thermodynamic trajectories with
peak densities above $2\times10^8\,\gccm$ in N100 (especially in its
deflagration phase), whereas in the violent merger all the burning
occurs at peak densities below $2\times10^8\,\gccm$. This leads to
vastly different degrees of neutronization in the ashes due to
electron capture reactions, resulting in higher abundances of stable
IGEs (in particular $^{54}$Fe, $^{56}$Fe, and $^{58}$Ni) for N100
which should be reflected in the presence of Ni lines in spectra from
the nebular phase \citep{maeda2010c,gerardy2007a}. Moreover, our
models differ significantly in composition and geometry of the
innermost ejecta. While the $^{56}$Ni distribution is roughly
spherical for N100, most of the $^{56}$Ni is off-center in the merger.
Asymmetric and shifted lines can be expected here and could correspond
to the effects discussed by \citet{maeda2010c, maeda2010b}. In the
merger, the detonation of the secondary WD at low densities produces
copious amounts of oxygen in the innermost ejecta. Potentially, this
could lead to strong [OI] $\lambda\lambda$6300,6364 emission at late
times \citep{kozma2005a}, which is not observed in SNe~Ia.  The
efficiency of [OI] emission, however, depends strongly on the
contamination with other elements and the ionization state of the
oxygen-rich zone. For strong ionization (as might be the case in
our low-density core), the presence of oxygen in the core would pose
no problem for the merger.  Clearly, nebular spectra contain important
information, but a firm conclusion can only be drawn upon detailed
three-dimensional modeling of the late-time RT, which is beyond the
scope of this letter.

There is, however, a more direct and perhaps more easily studied
effect.  Due to the low central densities of the two sub-Chandrasekhar
mass WDs, all of the IGEs in the merger are synthesized
in either $\alpha$-rich freeze-out from nuclear statistical
equilibrium or in incomplete Si-burning. In contrast, much of it is
produced under ``normal'' freeze-out conditions in the deflagration
phase of N100 (\citealt{thielemann1986a} place the dividing line
between $\alpha$-rich and ``normal'' freeze-out at
${\sim}2\times10^8\,\gccm$ for explosive burning of C+O material in
\sneia).  This leads to a higher abundance of $^{55}$Co---an isotope
mainly synthesized in the ``normal'' freeze-out and in incomplete
Si-burning (e.g.\ \citealt{thielemann1986a})---in the ejecta of the
Chandrasekhar-mass WD explosion than in those of the merger.

\begin{figure}
  \includegraphics[width=\linewidth]{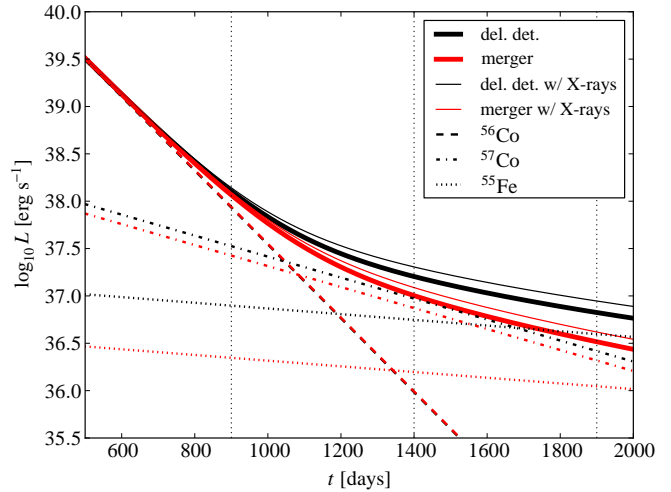}
  \caption{Leptonic luminosity as a function of time after the explosion 
    for N100 (black bold line) and the merger model (red bold line).
    The dashed, dash-dotted and dotted lines
    give the contribution to the leptonic luminosity due to the decay
    of selected isotopes. Thin solid lines give the luminosities
    including decay X-rays.\label{fig:latelc}}
\end{figure}

Such different isotopic ratios in the IGEs affect the shapes of
the predicted late-time light curves \citep{seitenzahl2009d}.
Starting at ${\sim}800\,\mathrm{d}$ after the explosion, the
\emph{leptonic light curves} that assume full transparency to
$\gamma$-rays and pure leptonic heating of the ejecta will be
increasingly powered by the decay of isotopes other than
$^{56}$Co. This is illustrated in Fig.~\ref{fig:latelc}. At
${\sim}1000\,\mathrm{d}$ after the explosion, the decay of $^{57}$Co to
$^{57}$Fe, which (in ${\sim}80\%$ of all decays) emits internal
conversion electrons, starts to dominate the light curves. Later, the
decay of $^{55}$Fe, which is mainly synthesized as $^{55}$Co, to
$^{55}$Mn (a ground-state to ground-state transition followed by the
emission of Auger electrons) contributes significantly and eventually
dominates the radioactive energy generation.

The leptonic light curve of N100  will fall off more
slowly than that of the merger. For example, the decrease in combined
leptonic energy production from $900\,\mathrm{d}$ to
$1400\,\mathrm{d}$ ($1900\,\mathrm{d}$) corresponds to a dimming by
2.25 (3.20) magnitudes for N100 and by 2.65 (3.87)
magnitudes for the merger (see
Table~\ref{tab:models}).

Thus, a measurement of the late light curve decline rate would
distinguish between an explosion of a Chandrasekhar-mass WD (which in
any scenario requires some pre-expansion in a deflagration stage) and
alternative models based on detonations in low density material---such
as mergers of WDs. For this, neither the correct distance to the
object nor the exact $^{56}$Ni production have to be known.  Note,
however, that the light curves shown are idealized cases assuming that
the leptonic energy production rates can be directly translated into
UVOIR light curves (which may be precluded by effects such as the
infrared catastrophe, ``frozen-in ionization'', CSM interaction,
leptonic losses, etc.).

\section{Conclusions}Although the nearby \snfe\ offers a unique
opportunity to scrutinize explosion models, at present a clear
preference of one scenario over the other is hard to establish. We
therefore discuss two models that are very distinct in the explosion
characteristics and in the resulting structure of the ejecta---a
delayed detonation of a Chandrasekhar-mass WD and a merger of two WDs
with a total of $2\,M_\odot$.

Comparing with early and near-maximum optical spectra, both
scenarios reproduce the main features but the merger is slightly
preferred because it provides a better match to the observations
around peak brightness. There are, however, shortcomings in other
aspects---such as the too long rise time---and therefore the working
hypothesis of \snfe\ resulting from a merger of two WDs requires
additional confirmation.

As shown here, alternatives to early-phase optical data may have
more decisive power. At very late epochs nucleosynthetic effects lead to different
characteristics in the photometric evolution that may allow us to
discriminate between explosion models. This, however, requires true
bolometric measurements, and it is unclear
at which wavelengths the maximum emission occurs at those late
epochs. Observations of \snfe\ will help to clarify this issue and
thus multi-wavelength monitoring of this object is essential. If the
maximum emission falls into the optical range, a clear distinction
between explosion models (that can then be related to progenitor
scenarios) will be possible from photometric measurements at
${\gtrsim}1000\,\mathrm{d}$. Thanks to its proximity, these observations should be feasible for
\snfe.\begin{acknowledgements}This work was supported by the Deutsche
  Forschungsgemeinschaft via the Transregional Collaborative Research
  Center TRR~33, the Emmy Noether Program (RO 3676/1-1) and the
  Excellence Cluster EXC~153, DOE Contracts DE-AC02-05CH1123 and
  DE-AC02-05CH11231, the Gordon \& Betty Moore Foundation, CNRS/IN2P3,
  CNRS/INSU, and PNC in France, the Max Planck Society, and the
  Tsinghua University Center for Astrophysics. The simulations were
  performed at JSC (grants PRACE042 and HMU014) and NCI at the ANU.\@
  We are grateful to C. Aspin, E.~Gaidos, A.~Mann, M.~Micheli,
  T.~Riesen, S.~Sonnett, and D.~Tholen, who granted us interrupt time
  to observe \snfe.\end{acknowledgements}


\begin{thebibliography}{44}
\expandafter\ifx\csname natexlab\endcsname\relax\def\natexlab#1{#1}\fi

\bibitem[{{Aldering} {et~al.}(2002){Aldering}, {Adam}, {Antilogus}, {Astier},
  {Bacon}, {Bongard}, {Bonnaud}, {Copin}, {Hardin}, {Henault}, {Howell},
  {Lemonnier}, {Levy}, {Loken}, {Nugent}, {Pain}, {Pecontal}, {Pecontal},
  {Perlmutter}, {Quimby}, {Schahmaneche}, {Smadja}, \&
  {Wood-Vasey}}]{aldering2002a}
{Aldering}, G., {Adam}, G., {Antilogus}, P., {Astier}, P., {Bacon}, R.,
  {Bongard}, S., {Bonnaud}, C., {Copin}, Y., {Hardin}, D., {Henault}, F.,
  {Howell}, D.~A., {Lemonnier}, J.-P., {Levy}, J.-M., {Loken}, S.~C., {Nugent},
  P.~E., {Pain}, R., {Pecontal}, A., {Pecontal}, E., {Perlmutter}, S.,
  {Quimby}, R.~M., {Schahmaneche}, K., {Smadja}, G., \& {Wood-Vasey}, W.~M.
  2002, in Society of Photo-Optical Instrumentation Engineers (SPIE) Conference
  Series, Vol. 4836, Survey and Other Telescope Technologies and Discoveries.,
  ed. J.~A. {Tyson} \& S.~{Wolff}, 61--72

\bibitem[{{Blondin} {et~al.}(2011){Blondin}, {Kasen}, {R{\"o}pke}, {Kirshner},
  \& {Mandel}}]{blondin2011a}
{Blondin}, S., {Kasen}, D., {R{\"o}pke}, F.~K., {Kirshner}, R.~P., \& {Mandel},
  K.~S. 2011, \mnras, 417, 1280

\bibitem[{{Bloom} {et~al.}(2012){Bloom}, {Kasen}, {Shen}, {Nugent}, {Butler},
  {Graham}, {Howell}, {Kolb}, {Holmes}, {Haswell}, {Burwitz}, {Rodriguez}, \&
  {Sullivan}}]{bloom2012a}
{Bloom}, J.~S., {Kasen}, D., {Shen}, K.~J., {Nugent}, P.~E., {Butler}, N.~R.,
  {Graham}, M.~L., {Howell}, D.~A., {Kolb}, U., {Holmes}, S., {Haswell}, C.~A.,
  {Burwitz}, V., {Rodriguez}, J., \& {Sullivan}, M. 2012, \apjl, 744, L17

\bibitem[{{Brown} {et~al.}(2011){Brown}, {Dawson}, {de Pasquale}, {Gronwall},
  {Holland}, {Immler}, {Kuin}, {Mazzali}, {Milne}, {Oates}, \&
  {Siegel}}]{brown2011a}
{Brown}, P.~J., {Dawson}, K.~S., {de Pasquale}, M., {Gronwall}, C., {Holland},
  S., {Immler}, S., {Kuin}, P., {Mazzali}, P., {Milne}, P., {Oates}, S., \&
  {Siegel}, M. 2011, ArXiv e-prints

\bibitem[{{Conley} {et~al.}(2006){Conley}, {Howell}, {Howes}, {Sullivan},
  {Astier}, {Balam}, {Basa}, {Carlberg}, {Fouchez}, {Guy}, {Hook}, {Neill},
  {Pain}, {Perrett}, {Pritchet}, {Regnault}, {Rich}, {Taillet}, {Aubourg},
  {Bronder}, {Ellis}, {Fabbro}, {Filiol}, {Le Borgne}, {Palanque-Delabrouille},
  {Perlmutter}, \& {Ripoche}}]{conley2006a}
{Conley}, A., {Howell}, D.~A., {Howes}, A., {Sullivan}, M., {Astier}, P.,
  {Balam}, D., {Basa}, S., {Carlberg}, R.~G., {Fouchez}, D., {Guy}, J., {Hook},
  I., {Neill}, J.~D., {Pain}, R., {Perrett}, K., {Pritchet}, C.~J., {Regnault},
  N., {Rich}, J., {Taillet}, R., {Aubourg}, E., {Bronder}, J., {Ellis}, R.~S.,
  {Fabbro}, S., {Filiol}, M., {Le Borgne}, D., {Palanque-Delabrouille}, N.,
  {Perlmutter}, S., \& {Ripoche}, P. 2006, \aj, 132, 1707

\bibitem[{{de Vaucouleurs} {et~al.}(1991){de Vaucouleurs}, {de Vaucouleurs},
  {Corwin}, {Buta}, {Paturel}, \& {Fouqu{\'e}}}]{devaucouleurs1991a}
{de Vaucouleurs}, G., {de Vaucouleurs}, A., {Corwin}, Jr., H.~G., {Buta},
  R.~J., {Paturel}, G., \& {Fouqu{\'e}}, P. 1991, {Third Reference Catalogue of
  Bright Galaxies} (Berlin Heidelberg New York: Springer-Verlag)

\bibitem[{{Gerardy} {et~al.}(2007){Gerardy}, {Meikle}, {Kotak}, {H{\"o}flich},
  {Farrah}, {Filippenko}, {Foley}, {Lundqvist}, {Mattila}, {Pozzo},
  {Sollerman}, {Van Dyk}, \& {Wheeler}}]{gerardy2007a}
{Gerardy}, C.~L., {Meikle}, W.~P.~S., {Kotak}, R., {H{\"o}flich}, P., {Farrah},
  D., {Filippenko}, A.~V., {Foley}, R.~J., {Lundqvist}, P., {Mattila}, S.,
  {Pozzo}, M., {Sollerman}, J., {Van Dyk}, S.~D., \& {Wheeler}, J.~C. 2007,
  \apj, 661, 995

\bibitem[{{Hayden} {et~al.}(2010){Hayden}, {Garnavich}, {Kessler}, {Frieman},
  {Jha}, {Bassett}, {Cinabro}, {Dilday}, {Kasen}, {Marriner}, {Nichol},
  {Riess}, {Sako}, {Schneider}, {Smith}, \& {Sollerman}}]{hayden2010a}
{Hayden}, B.~T., {Garnavich}, P.~M., {Kessler}, R., {Frieman}, J.~A., {Jha},
  S.~W., {Bassett}, B., {Cinabro}, D., {Dilday}, B., {Kasen}, D., {Marriner},
  J., {Nichol}, R.~C., {Riess}, A.~G., {Sako}, M., {Schneider}, D.~P., {Smith},
  M., \& {Sollerman}, J. 2010, \apj, 712, 350

\bibitem[{{Kasen} {et~al.}(2009){Kasen}, {R{\"o}pke}, \&
  {Woosley}}]{kasen2009a}
{Kasen}, D., {R{\"o}pke}, F.~K., \& {Woosley}, S.~E. 2009, \nat, 460, 869

\bibitem[{{Khokhlov}(1991)}]{khokhlov1991a}
{Khokhlov}, A.~M. 1991, \aap, 245, 114

\bibitem[{{Kozma} {et~al.}(2005){Kozma}, {Fransson}, {Hillebrandt},
  {Travaglio}, {Sollerman}, {Reinecke}, {R{\"o}pke}, \&
  {Spyromilio}}]{kozma2005a}
{Kozma}, C., {Fransson}, C., {Hillebrandt}, W., {Travaglio}, C., {Sollerman},
  J., {Reinecke}, M., {R{\"o}pke}, F.~K., \& {Spyromilio}, J. 2005, \aap, 437,
  983

\bibitem[{{Kromer} \& {Sim}(2009)}]{kromer2009a}
{Kromer}, M., \& {Sim}, S.~A. 2009, \mnras, 398, 1809

\bibitem[{Kurucz \& Bell(1995)}]{kurucz1995a}
Kurucz, R., \& Bell, B. 1995, {Cambridge, Mass.: Smithsonian Astrophysical
  Observatory.}

\bibitem[{{Li} {et~al.}(2011){Li}, {Bloom}, {Podsiadlowski}, {Miller}, {Cenko},
  {Jha}, {Sullivan}, {Howell}, {Nugent}, {Butler}, {Ofek}, {Kasliwal},
  {Richards}, {Stockton}, {Shih}, {Bildsten}, {Shara}, {Bibby}, {Filippenko},
  {Ganeshalingam}, {Silverman}, {Kulkarni}, {Law}, {Poznanski}, {Quimby},
  {McCully}, {Patel}, {Maguire}, \& {Shen}}]{li2011b}
{Li}, W., {Bloom}, J.~S., {Podsiadlowski}, P., {Miller}, A.~A., {Cenko}, S.~B.,
  {Jha}, S.~W., {Sullivan}, M., {Howell}, D.~A., {Nugent}, P.~E., {Butler},
  N.~R., {Ofek}, E.~O., {Kasliwal}, M.~M., {Richards}, J.~W., {Stockton}, A.,
  {Shih}, H.-Y., {Bildsten}, L., {Shara}, M.~M., {Bibby}, J., {Filippenko},
  A.~V., {Ganeshalingam}, M., {Silverman}, J.~M., {Kulkarni}, S.~R., {Law},
  N.~M., {Poznanski}, D., {Quimby}, R.~M., {McCully}, C., {Patel}, B.,
  {Maguire}, K., \& {Shen}, K.~J. 2011, \nat, 480, 348

\bibitem[{{Liu} {et~al.}(2011){Liu}, {Di Stefano}, {Wang}, \& {Moe}}]{liu2011a}
{Liu}, J., {Di Stefano}, R., {Wang}, T., \& {Moe}, M. 2011, ArXiv e-prints

\bibitem[{{Maeda} {et~al.}(2010{\natexlab{a}}){Maeda}, {Benetti},
  {Stritzinger}, {R{\"o}pke}, {Folatelli}, {Sollerman}, {Taubenberger},
  {Nomoto}, {Leloudas}, {Hamuy}, {Tanaka}, {Mazzali}, \&
  {Elias-Rosa}}]{maeda2010b}
{Maeda}, K., {Benetti}, S., {Stritzinger}, M., {R{\"o}pke}, F.~K., {Folatelli},
  G., {Sollerman}, J., {Taubenberger}, S., {Nomoto}, K., {Leloudas}, G.,
  {Hamuy}, M., {Tanaka}, M., {Mazzali}, P.~A., \& {Elias-Rosa}, N.
  2010{\natexlab{a}}, \nat, 466, 82

\bibitem[{{Maeda} {et~al.}(2010{\natexlab{b}}){Maeda}, {Taubenberger},
  {Sollerman}, {Mazzali}, {Leloudas}, {Nomoto}, \& {Motohara}}]{maeda2010c}
{Maeda}, K., {Taubenberger}, S., {Sollerman}, J., {Mazzali}, P.~A., {Leloudas},
  G., {Nomoto}, K., \& {Motohara}, K. 2010{\natexlab{b}}, \apj, 708, 1703

\bibitem[{{Mazzali} {et~al.}(2007){Mazzali}, {R{\"o}pke}, {Benetti}, \&
  {Hillebrandt}}]{mazzali2007a}
{Mazzali}, P.~A., {R{\"o}pke}, F.~K., {Benetti}, S., \& {Hillebrandt}, W. 2007,
  Science, 315, 825

\bibitem[{{Mazzali} {et~al.}(2008){Mazzali}, {Sauer}, {Pastorello}, {Benetti},
  \& {Hillebrandt}}]{mazzali2008b}
{Mazzali}, P.~A., {Sauer}, D.~N., {Pastorello}, A., {Benetti}, S., \&
  {Hillebrandt}, W. 2008, \mnras, 386, 1897

\bibitem[{{Nomoto} {et~al.}(1984){Nomoto}, {Thielemann}, \&
  {Yokoi}}]{nomoto1984a}
{Nomoto}, K., {Thielemann}, F.-K., \& {Yokoi}, K. 1984, \apj, 286, 644

\bibitem[{{Nugent} {et~al.}(2011{\natexlab{a}}){Nugent}, {Sullivan}, {Bersier},
  {Howell}, {Thomas}, \& {James}}]{nugent2011b}
{Nugent}, P., {Sullivan}, M., {Bersier}, D., {Howell}, D.~A., {Thomas}, R., \&
  {James}, P. 2011{\natexlab{a}}, The Astronomer's Telegram, 3581, 1

\bibitem[{{Nugent} {et~al.}(2011{\natexlab{b}}){Nugent}, {Sullivan}, {Cenko},
  {Thomas}, {Kasen}, {Howell}, {Bersier}, {Bloom}, {Kulkarni}, {Kandrashoff},
  {Filippenko}, {Silverman}, {Marcy}, {Howard}, {Isaacson}, {Maguire},
  {Suzuki}, {Tarlton}, {Pan}, {Bildsten}, {Fulton}, {Parrent}, {Sand},
  {Podsiadlowski}, {Bianco}, {Dilday}, {Graham}, {Lyman}, {James}, {Kasliwal},
  {Law}, {Quimby}, {Hook}, {Walker}, {Mazzali}, {Pian}, {Ofek}, {Gal-Yam}, \&
  {Poznanski}}]{nugent2011a}
{Nugent}, P.~E., {Sullivan}, M., {Cenko}, S.~B., {Thomas}, R.~C., {Kasen}, D.,
  {Howell}, D.~A., {Bersier}, D., {Bloom}, J.~S., {Kulkarni}, S.~R.,
  {Kandrashoff}, M.~T., {Filippenko}, A.~V., {Silverman}, J.~M., {Marcy},
  G.~W., {Howard}, A.~W., {Isaacson}, H.~T., {Maguire}, K., {Suzuki}, N.,
  {Tarlton}, J.~E., {Pan}, Y.-C., {Bildsten}, L., {Fulton}, B.~J., {Parrent},
  J.~T., {Sand}, D., {Podsiadlowski}, P., {Bianco}, F.~B., {Dilday}, B.,
  {Graham}, M.~L., {Lyman}, J., {James}, P., {Kasliwal}, M.~M., {Law}, N.~M.,
  {Quimby}, R.~M., {Hook}, I.~M., {Walker}, E.~S., {Mazzali}, P., {Pian}, E.,
  {Ofek}, E.~O., {Gal-Yam}, A., \& {Poznanski}, D. 2011{\natexlab{b}}, \nat,
  480, 344

\bibitem[{{Pakmor} {et~al.}(2012){Pakmor}, {Kromer}, {Taubenberger}, {Sim},
  {R{\"o}pke}, \& {Hillebrandt}}]{pakmor2012a}
{Pakmor}, R., {Kromer}, M., {Taubenberger}, S., {Sim}, S.~A., {R{\"o}pke},
  F.~K., \& {Hillebrandt}, W. 2012, \apjl, 747, L10

\bibitem[{{Reinecke} {et~al.}(1999){Reinecke}, {Hillebrandt}, {Niemeyer},
  {Klein}, \& {Gr{\"o}bl}}]{reinecke1999a}
{Reinecke}, M., {Hillebrandt}, W., {Niemeyer}, J.~C., {Klein}, R., \&
  {Gr{\"o}bl}, A. 1999, \aap, 347, 724

\bibitem[{{R{\"o}pke}(2005)}]{roepke2005c}
{R{\"o}pke}, F.~K. 2005, \aap, 432, 969

\bibitem[{{R{\"o}pke}(2007)}]{roepke2007d}
---. 2007, \apj, 668, 1103

\bibitem[{{R{\"o}pke} \& {Hillebrandt}(2005)}]{roepke2005b}
{R{\"o}pke}, F.~K., \& {Hillebrandt}, W. 2005, \aap, 431, 635

\bibitem[{{R{\"o}pke} {et~al.}(2007){R{\"o}pke}, {Hillebrandt}, {Schmidt},
  {Niemeyer}, {Blinnikov}, \& {Mazzali}}]{roepke2007c}
{R{\"o}pke}, F.~K., {Hillebrandt}, W., {Schmidt}, W., {Niemeyer}, J.~C.,
  {Blinnikov}, S.~I., \& {Mazzali}, P.~A. 2007, \apj, 668, 1132

\bibitem[{{R{\"o}pke} \& {Niemeyer}(2007)}]{roepke2007b}
{R{\"o}pke}, F.~K., \& {Niemeyer}, J.~C. 2007, \aap, 464, 683

\bibitem[{{Schlegel} {et~al.}(1998){Schlegel}, {Finkbeiner}, \&
  {Davis}}]{schlegel1998a}
{Schlegel}, D.~J., {Finkbeiner}, D.~P., \& {Davis}, M. 1998, \apj, 500, 525

\bibitem[{{Schmidt} {et~al.}(2006){Schmidt}, {Niemeyer}, {Hillebrandt}, \&
  {R{\"o}pke}}]{schmidt2006c}
{Schmidt}, W., {Niemeyer}, J.~C., {Hillebrandt}, W., \& {R{\"o}pke}, F.~K.
  2006, \aap, 450, 283

\bibitem[{{Seitenzahl} {et~al.}(2009{\natexlab{a}}){Seitenzahl}, {Meakin},
  {Townsley}, {Lamb}, \& {Truran}}]{seitenzahl2009b}
{Seitenzahl}, I.~R., {Meakin}, C.~A., {Townsley}, D.~M., {Lamb}, D.~Q., \&
  {Truran}, J.~W. 2009{\natexlab{a}}, \apj, 696, 515

\bibitem[{{Seitenzahl} {et~al.}(2010){Seitenzahl}, {R{\"o}pke}, {Fink}, \&
  {Pakmor}}]{seitenzahl2010a}
{Seitenzahl}, I.~R., {R{\"o}pke}, F.~K., {Fink}, M., \& {Pakmor}, R. 2010,
  \mnras, 407, 2297

\bibitem[{{Seitenzahl} {et~al.}(2009{\natexlab{b}}){Seitenzahl},
  {Taubenberger}, \& {Sim}}]{seitenzahl2009d}
{Seitenzahl}, I.~R., {Taubenberger}, S., \& {Sim}, S.~A. 2009{\natexlab{b}},
  \mnras, 400, 531

\bibitem[{{Shappee} \& {Stanek}(2011)}]{shappee2011a}
{Shappee}, B.~J., \& {Stanek}, K.~Z. 2011, \apj, 733, 124

\bibitem[{{Sim}(2007)}]{sim2007b}
{Sim}, S.~A. 2007, \mnras, 375, 154

\bibitem[{{Smith} {et~al.}(2011){Smith}, {Williams}, {Smith}, {Milne},
  {Jannuzi}, \& {Green}}]{smith2011a}
{Smith}, P.~S., {Williams}, G.~G., {Smith}, N., {Milne}, P.~A., {Jannuzi},
  B.~T., \& {Green}, E.~M. 2011, ArXiv e-prints

\bibitem[{{Springel}(2005)}]{springel2005a}
{Springel}, V. 2005, \mnras, 364, 1105

\bibitem[{{Stehle} {et~al.}(2005){Stehle}, {Mazzali}, {Benetti}, \&
  {Hillebrandt}}]{stehle2005a}
{Stehle}, M., {Mazzali}, P.~A., {Benetti}, S., \& {Hillebrandt}, W. 2005,
  \mnras, 360, 1231

\bibitem[{{Stritzinger} {et~al.}(2006){Stritzinger}, {Leibundgut}, {Walch}, \&
  {Contardo}}]{stritzinger2006a}
{Stritzinger}, M., {Leibundgut}, B., {Walch}, S., \& {Contardo}, G. 2006, \aap,
  450, 241

\bibitem[{{Thielemann} {et~al.}(1986){Thielemann}, {Nomoto}, \&
  {Yokoi}}]{thielemann1986a}
{Thielemann}, F.-K., {Nomoto}, K., \& {Yokoi}, K. 1986, \aap, 158, 17

\bibitem[{{Travaglio} {et~al.}(2004){Travaglio}, {Hillebrandt}, {Reinecke}, \&
  {Thielemann}}]{travaglio2004a}
{Travaglio}, C., {Hillebrandt}, W., {Reinecke}, M., \& {Thielemann}, F.-K.
  2004, \aap, 425, 1029

\bibitem[{{Wang} \& {Wheeler}(2008)}]{wang2008a}
{Wang}, L., \& {Wheeler}, J.~C. 2008, \araa, 46, 433

\bibitem[{{Woosley} {et~al.}(2009){Woosley}, {Kerstein}, {Sankaran}, {Aspden},
  \& {R{\"o}pke}}]{woosley2009a}
{Woosley}, S.~E., {Kerstein}, A.~R., {Sankaran}, V., {Aspden}, A.~J., \&
  {R{\"o}pke}, F.~K. 2009, \apj, 704, 255

\end{thebibliography}
\end{document}